\documentclass[sigconf]{acmart}

\AtBeginDocument{%
  \providecommand\BibTeX{{%
    \normalfont B\kern-0.5em{\scshape i\kern-0.25em b}\kern-0.8em\TeX}}}

\usepackage[utf8]{inputenc} 
\usepackage[T1]{fontenc}    
\usepackage{hyperref}       
\usepackage{url}            
\usepackage{booktabs}       
\usepackage{amsfonts}       
\usepackage{nicefrac}       
\usepackage{microtype}      
\usepackage{xcolor}         
\usepackage{multirow}
\usepackage{graphicx}
\usepackage{subcaption}
\usepackage[algo2e]{algorithm2e} 

\usepackage{algorithm}  
\usepackage{algorithmicx}  
\usepackage{algpseudocode} 
\usepackage{comment}
\usepackage{graphicx}
\usepackage{subcaption}

\newcommand{\red}[1]{{\color{red} {#1}}}

\setcopyright{none}
\renewcommand\footnotetextcopyrightpermission[1]{}
\begin{document}

\title{Automating Energy-Efficient GPU Kernel Generation: A Fast Search-Based Compilation Approach}

\author{Yijia Zhang$^1$, Zhihong Gou$^1$, Shijie Cao$^2$, Weigang Feng$^1$, Sicheng Zhang$^1$,}
\author{Guohao Dai$^1$, Ningyi Xu$^1$}
\authornote{Corresponding author.}
\email{{zhangyijia, gouzhihong91, fengweigang, zhangsicheng, daiguohao, xuningyi}@sjtu.edu.cn}
\email{shijiecao@microsoft.com}
\affiliation{%
  \institution{$^1$Shanghai Jiao Tong University, $^2$Microsoft Research Asia}
  \country{}
}

\begin{abstract}

Deep Neural Networks (DNNs) have revolutionized various fields, but their deployment on GPUs often leads to significant energy consumption.
Unlike existing methods for reducing GPU energy consumption, which are either hardware-inflexible or limited by workload constraints, this paper addresses the problem at the GPU kernel level.
We propose a novel search-based compilation method to generate energy-efficient GPU kernels by incorporating energy efficiency into the search process.
To accelerate the energy evaluation process, we develop an accurate energy cost model based on high-level kernel features.
Furthermore, we introduce a dynamic updating strategy for the energy cost model, reducing the need for on-device energy measurements and accelerating the search process.
Our evaluation demonstrates that the proposed approach can generate GPU kernels with up to 21.69\% reduced energy consumption while maintaining low latency.
\end{abstract}


\maketitle

\section{Introduction}

Deep Neural Networks (DNNs) have demonstrated outstanding performance in various fields, including computer vision~\cite{deng2009imagenet, he2016deep}, natural language processing~\cite{kenton2019bert}, and generative artificial intelligence (AI)\cite{jiang2022vima}.
Consequently, many vendors have trained or deployed DNNs in GPU cloud environments to generate profits\cite{yu2022orca}. For instance, Meta engineers trained LLaMA-3~\cite{LLaMA3} on a computer cluster comprising 24,576 NVIDIA H100 GPUs. To further explore the potential of artificial intelligence, Meta plans to scale their cluster to 350,000 H100 GPUs~\cite{Inf}.

During the training or deployment of Deep Neural Networks (DNNs), one significant challenge is the increasing energy consumption. The electrical energy required for the largest AI training runs has been increasing exponentially, with a 3.4-month doubling time observed from AlexNet to AlphaGo Zero~\cite{openai2018}.
Currently, with the widespread adoption of generative AI, energy consumption is expected to rise even further~\cite{aineedsmoreelec}. For example, training the GPT-3 model~\cite{brown2020language} consumes 1,287 megawatt-hours (MWh) of electricity~\cite{patterson2021carbon}, equivalent to 120 years of electricity consumption for an average U.S. household.
Beyond the enormous electricity costs, the sudden surge in demand from GPU clusters can overwhelm power supply systems. For instance, OpenAI has reportedly projected a demand of over 100 MW for future GPT training~\cite{damagepowergrid}, a requirement that could exceed the capacity of a single state’s power grid without causing disruptions.
Thus, addressing energy reduction in GPU clusters is imperative.

The energy consumed by a GPU cluster can typically be divided into two categories~\cite{dayarathna2015data}: energy used by IT equipment (e.g., GPU servers, networks, storage, etc.) and energy used by infrastructure facilities (e.g., cooling and power conditioning systems).
Notably, the operating power of the servers directly impacts the energy consumption of infrastructure facilities. For example, the power required to run an air-cooling system is cubically proportional to the servers’ operating power.
Since cooling-related energy expenses account for approximately 50\% of a typical cluster’s total energy consumption~\cite{dayarathna2015data}, reducing the operating power of GPU servers not only lowers their energy costs but also significantly contributes to reducing overall cluster expenses.

To reduce the operating power and energy consumption of GPUs, several methods have been proposed, including chip-level, workload-level, and kernel-level approaches.
For chip-level methods, GPU manufacturers provide features such as GPU power capping~\cite{powerlimitting}, which restricts GPU operating power.
In addition, modern GPU chips support manual voltage and frequency adjustment, which can also be leveraged to manage operating power. Based on these chip features, several studies have proposed strategies to reduce GPU energy consumption~\cite{wang2021dynamic, bharadwaj2021dub, zou2020indicator}. While effective, these methods may introduce system instability due to manual adjustments of chip properties.
For workload-level methods, researchers have developed an optimization framework named Zeus~\cite{you2023zeus}, which automatically determines the optimal batch size to minimize energy consumption during DNN training. However, this approach lacks flexibility in scenarios where batch size specification is constrained.
For kernel-level methods, Jayaweera et al.~\cite{jayaweera2024energy} proposed a novel tile size selection strategy that balances the trade-off between data reuse and data sharing, thereby improving both performance and energy efficiency. Nonetheless, the limited exploration space for kernels in this approach may result in sub-optimal kernel implementations.

To explore the full design space of tensor programs, we propose a search-based compilation method for energy-efficient kernel generation.
Currently, various search-based compilation methods (e.g., TVM~\cite{chen2018tvm}, Ansor~\cite{zheng2020ansor}) have been developed to produce high-performance kernels.
Compared with vendor-provided kernels, such as those in cuDNN~\cite{chetlur2014cudnn}, kernels generated through search-based compilation methods offer better flexibility and comparable performance.
However, existing search-based kernel generation methods primarily focus on kernel latency while neglecting energy efficiency, which can result in kernels with low latency but poor energy efficiency.
Given the critical energy challenges in GPU clusters, it is increasingly important to incorporate power and energy considerations into search-based kernel generation methods.

Achieving this goal presents several challenges.
First, reducing operating power is not without cost: the reduction often exhibits a non-linear relationship with increased execution latency, which can worsen overall energy consumption.
Second, using on-device measurements for energy evaluation is time-intensive, necessitating the development of an energy cost model for kernels.
Finally, introducing a cost-model-based energy evaluation may compromise the quality of the searched kernels.

\begin{table}[h]
\caption{Compared with related methods, our method is a search-based compilation approach with fast energy evaluation for energy-efficient kernel generation.}
\resizebox{0.8\columnwidth}{!}{%
\begin{tabular}{c|cccc}
\hline
 & ODPP~\cite{zou2020indicator} & Zeus~\cite{you2023zeus} &  Ansor~\cite{zheng2020ansor} & Ours \\ \hline
Energy aware & $\checkmark$ & $\checkmark$  &  & $\checkmark$ \\
System flexible &  & $\checkmark$  & $\checkmark$ & $\checkmark$ \\
Workload friendly & $\checkmark$ &   & $\checkmark$ & $\checkmark$ \\
Big exploration space &  & $\checkmark$   & $\checkmark$ & $\checkmark$ \\
Fast energy evaluation & $\checkmark$ &    &  & $\checkmark$ \\ \hline
\end{tabular}%
}

\label{tab:compare} 
\end{table}

To address these challenges, we propose a novel search-based method for generating kernels with high performance and energy efficiency, which distinguishes itself from related works summarized in Table~\ref{tab:compare}.
Our contributions are as follows:

\begin{enumerate}
    \item To the best of our knowledge, we develop the first search-based energy-aware GPU kernel generation framework.
    \item We propose an efficient energy cost model to predict kernel energy consumption, significantly accelerating the energy evaluation process.
    \item We introduce a dynamic online updating strategy for the energy cost model, which speeds up the search process while preserving search quality.
\end{enumerate}

The remainder of this paper is organized as follows. Section~\ref{sec:background} provides the background of this work. Section~\ref{sec:overview} presents an overview of the proposed energy-aware GPU kernel generation framework. Subsequently, the framework’s details are discussed, including the energy-aware search process, energy cost model, and online updating strategy, which are covered in Sections~\ref{sec:pipeline}, \ref{sec:costmodel}, and \ref{sec:strategy}, respectively.
Experimental results are presented and analyzed in Section\ref{sec:exp}. To further elucidate the energy characteristics of kernels, a case study is conducted in Section\ref{sec:casestudy}. Finally, conclusions are drawn in Section~\ref{sec:conclusion}.

\section{Background} ~\label{sec:background}

\subsection{GPU architecture}
In typical GPU architectures, the Streaming Processor (SP) is the smallest processing unit and is referred to as CUDA cores in Nvidia GPUs. A Streaming MultiProcessor (SM) consists of multiple SPs, with the exact number varying across different architectures. For example, in Nvidia Pascal GPUs, one SM comprises 128 SPs.
Before a kernel is launched on the GPU, it must be divided into numerous thread blocks and allocated across different SMs for execution. During execution, threads share the limited memory resources of their respective SMs, including registers and shared memory.
The GPU’s memory hierarchy further includes L1 cache, L2 cache, and global memory, all designed to improve the kernel’s data reuse rate and thereby accelerate execution.

\subsection{Deep Learning Compilers}
To generate efficient kernels for GPUs, deep learning compilers need to map computational programs using GPU intrinsics and perform optimization passes on them. To achieve better optimization for kernels, Halide~\cite{ragan2013halide} introduced the methodology of compute and schedule, with the former describing the computational logic and the latter specifying the optimization methods. Some existing compilers, including AutoTVM~\cite{chen2018learning} and UNIT~\cite{weng2021unit}, rely on hand-written templates to specify the schedule, while others can find schedules automatically, such as FlexTensor~\cite{zheng2020flextensor}, Rammer~\cite{ma2020rammer}, and Ansor~\cite{zheng2020ansor}. These automatic schedulers search for efficient schedules in a manually designed search space and optimize the kernel implementations. For example, Ansor samples kernel implementations from a hierarchical representation of the kernel search space and then fine-tunes the kernels using evolutionary search and a learned latency cost model built on XGBoost~\cite{chen2016xgboost}.

\subsection{Decomposition of GPU Energy Consumption} 
The energy consumption of a GPU is equal to the average power of the GPU during operation multiplied by the duration of the GPU runtime. GPU energy consumption can be divided into three parts~\cite{kandiah2021accelwattch}: constant power, static power, and dynamic power. Constant power can be generated by board fans and peripheral circuits. Static power appears once hardware components are activated by turning on the circuit power gates, even without the logic reversal of transistors. Research reveals that constant and static power account for an average of 40-50\% of GPU power across different GPUs. When memory and computation components are executed by GPU programs, GPUs consume dynamic power in addition to constant power and static power. Usually, the energy consumed by memory access can account for more than half of the dynamic power.
\begin{figure}[h]  
    \centering  
        \includegraphics[width=0.45\textwidth]{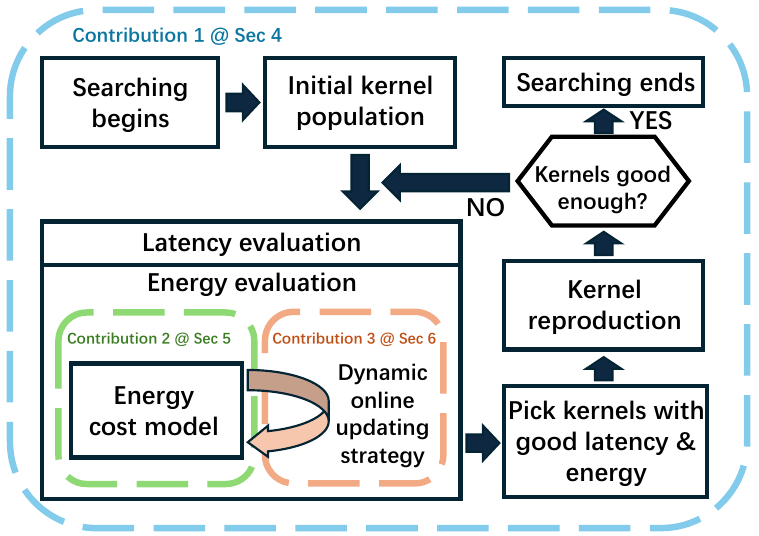}  
    \caption{Our search-based compilation for energy-efficient kernel generation.} 
    \label{fig:overview} 
\end{figure}

\section{overview} ~\label{sec:overview}
In this paper, we propose a method for generating energy-efficient kernels for deep learning, as shown in Figure~\ref{fig:overview}. Unlike previous auto-schedulers that solely focus on searching for kernels with optimal latency, we incorporate kernel energy considerations into the search process for the first time. By using genetic algorithms, we can find kernels with good latency and lower energy consumption by setting both latency and energy as reproduction genes. Furthermore, we endeavor to accelerate kernel energy evaluation by designing a machine-learning-based energy cost model, which speeds up the energy evaluation of a kernel by thousands of times compared to using the GPU’s built-in energy measurement APIs. Additionally, to reduce the number of energy measurements while maintaining accuracy, we propose a dynamic updating strategy for the energy cost model, thereby accelerating the search process.

\begin{figure}[h]  
    \centering  
        \includegraphics[width=0.45\textwidth]{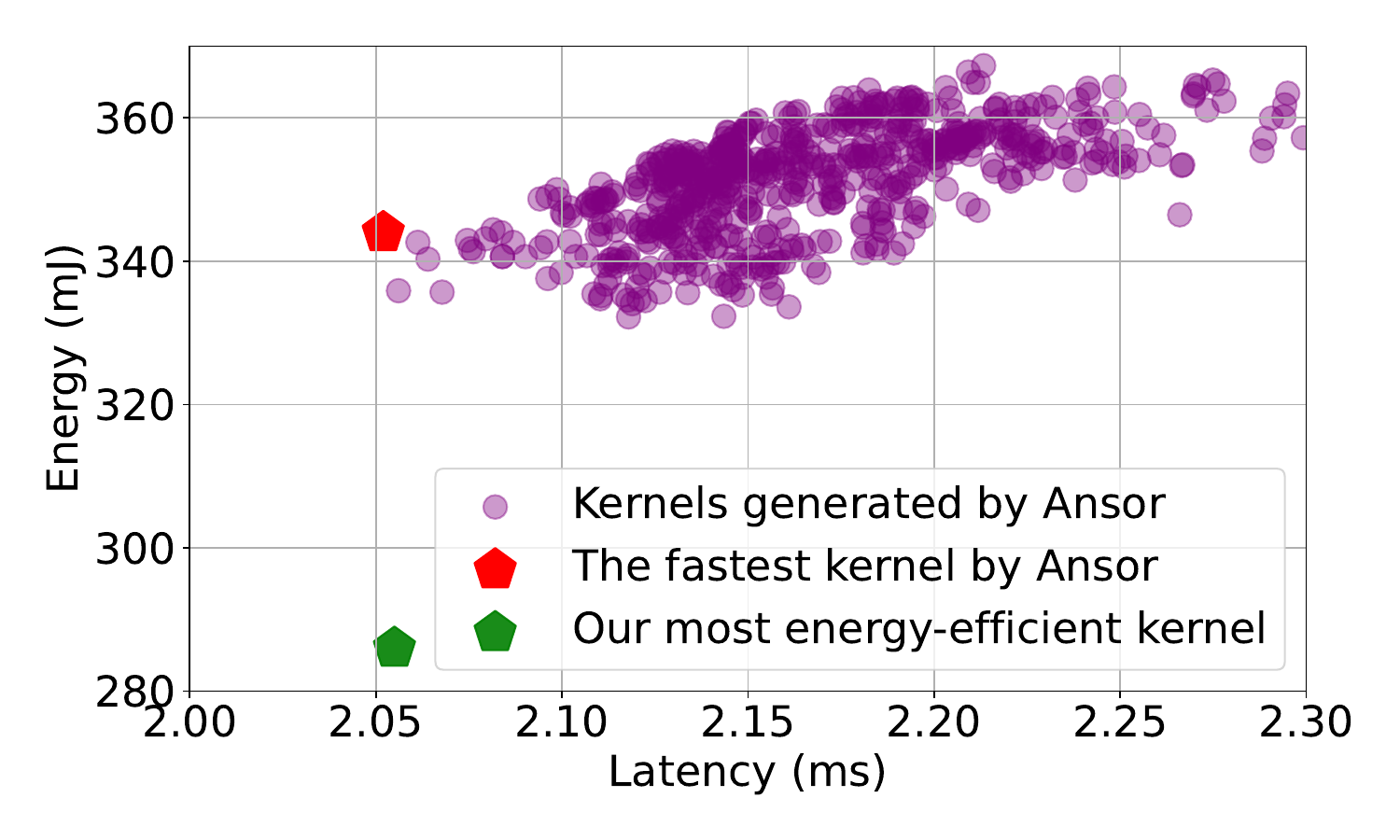}  
    \caption{The latency and energy consumption of one convolution operator from ResNet-50 generated by Ansor, running on one NVIDIA P100 GPU. The kernel generate by our method consumes less energy while maintaining similar latency with Ansor's.} 
    \label{fig:ansor_kernels} 
\end{figure} 

\section{Energy-efficient kernel generation method} ~\label{sec:pipeline}
In this section, we will introduce our energy-efficient auto-scheduled kernel generation method, which can generate efficient kernels with both low latency and low energy consumption.

\subsection{Motivation}
Typically, it is assumed that kernels with lower latency also consume less energy, based on the assumption that various kernels operate at similar average power levels. However, our findings challenge this assumption. Figure~\ref{fig:ansor_kernels} illustrates the energy consumption of the convolution (Conv) kernels generated through Ansor's search process. It can be observed that even for the same operator, the energy consumed by different kernel implementations can vary significantly. In some instances, kernels with comparable latency exhibit notable differences in energy consumption. Additionally, we discover that among the kernels generated by Ansor, some exhibit latency close to that of the most latency-efficient kernel yet consume less power, resulting in reduced energy consumption. Consequently, we posit that incorporating energy metrics into the kernel search process has the potential to identify kernels that are superior in energy efficiency.

\subsection{Challenge}
\begin{figure}[h]  
    \centering  
        \includegraphics[width=0.4\textwidth]{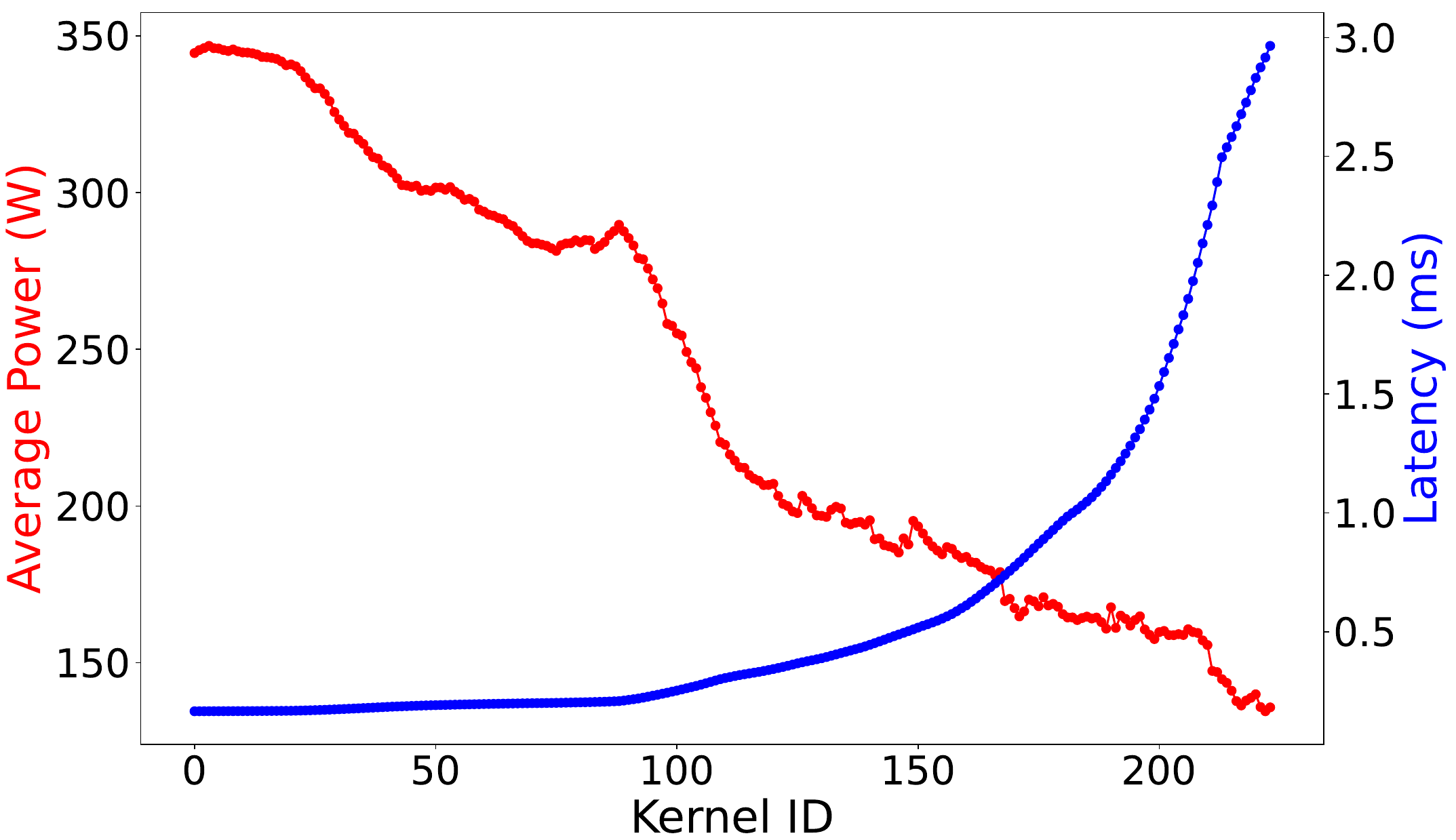}  
    \caption{The inverse correlation between latency and operating power of MatMul (M, N, K=1024, 1024, 1024) kernels generated by Ansor. Evaluation has been conducted on one NVIDIA A100 GPU.} 
    \label{fig:sortedlatvspwr} 
\end{figure}

We have observed a certain degree of inverse correlation between kernel latency and operating power. As illustrated in Figure~\ref{fig:sortedlatvspwr}, there is a trend where higher latency in the MatMul kernel coincides with a decrease in its average power usage. Considering that the energy consumption of a kernel is equal to the product of its latency and average power, pinpointing the kernel with the lowest energy expenditure is not straightforward. Furthermore, kernels with higher latency often do not comply with the strict timing constraints required in various scenarios. Consequently, it is essential to discover strategies that enable the reduction of energy consumption without compromising the kernel's latency.

\subsection{Insight}
An analysis of Figure~\ref{fig:ansor_kernels} reveals that kernels with higher latency tend to have greater energy consumption because energy is the product of kernel latency and average power. Therefore, even if a kernel has lower average power, it can still have higher energy consumption if it runs slowly. This insight highlights the importance of lower latency for energy reduction.
We incorporate this understanding into our search method. Specifically, we employ a genetic algorithm for searching and select kernels with lower latency at each reproduction iteration. Subsequently, we choose the most energy-efficient kernels among these low-latency candidates for the next stage of evolution.

\subsection{Approach}
To generate energy-efficient kernels, both accurate energy evaluation and an effective generation method are important. We will detail our method for measuring kernel power and energy, followed by an explanation of our energy-efficient kernel generation method.

\textbf{Power and energy Measurement.} To obtain GPU energy values during kernel execution, we establish a kernel power measurement framework based on NVIDIA Management Library (NVML), NVIDIA's power measurement API~\cite{nvml}. Before launching the kernel on the GPU, we run a pre-heating kernel for several seconds to warm up the GPU. This step is crucial because studies have shown that a GPU's operating temperature significantly affects energy consumption~\cite{hong2010integrated}. By pre-heating, we ensure the GPU temperature remains consistent for each measurement.
To measure kernel power accurately, we execute the kernel repetitively for thousands of iterations to minimize power measurement errors. During execution, the NVML API samples the GPU's power at the highest sampling rate it supports. Once the kernel finishes executing, we calculate the average power during runtime by averaging the power samples over time. The energy consumed during a single run of the kernel is then determined by multiplying the average power by the latency of a single run.

\textbf{Energy-efficient kernel generation method.} We employ a genetic algorithm to produce energy-efficient kernels. During each iteration of the genetic algorithm, new kernels are generated. To ensure that these kernels maintain optimal latency while reducing energy consumption, we start by selecting kernels with the lowest latency from the generation. We then evaluate the energy consumption of these faster kernels and select those with lower energy usage. These kernels, which exhibit both good latency and low energy consumption, are used to reproduce the next generation of kernels. This search process is repeated until the results converge.

\section{Energy cost model} ~\label{sec:costmodel}
In this chapter, we introduce our energy cost model, which we use to accelerate energy evaluation during kernel search.

\subsection{Motivation}
In the power and energy measurement approach in Section~\ref{sec:pipeline}, we use the NVML API to obtain the energy consumption of each candidate kernel. However, this measurement process is time-consuming due to several factors:

\begin{enumerate}
    \item The NVML API is constrained by a sampling frequency of 30 to 50 Hz, whereas the execution time of a kernel is often on the order of milliseconds—much shorter than the sampling period. To achieve accurate power measurements, each kernel must execute thousands of times to obtain the average power values sampled through the NVML API, which could take several seconds.
    \item Temperature variations significantly affect transistor behavior, leading to notable differences in GPU energy consumption even when executing the same workload. To ensure reliable energy data, each kernel measurement is preceded by a warm-up period of several seconds to stabilize the GPU at a consistent temperature.
\end{enumerate}

Therefore, we need to accelerate the energy evaluation process for better searching efficiency.

\subsection{Challenge}
In addition to using the NVML API, alternative methods for estimating GPU power have been developed based on the behavior of the GPU during kernel execution. These methods are characterized by relatively accurate power predictions but remain time-consuming. A notable example is presented in AccelWattch~\cite{kandiah2021accelwattch}, which designs a power prediction model for modern GPUs. This model relies on hardware counters from different GPU modules during kernel execution. To obtain these counters, time-intensive tools are used, including GPGPU-Sim~\cite{lew2019analyzing}, a GPU behavior simulator, and NVProf, a GPU profiling tool with built-in API capabilities. Gathering these counters can take from a few seconds to several hours per kernel. Furthermore, the time-consuming nature of current GPU kernel power measurement methods results in limited data collection. Therefore, accurately and quickly predicting GPU power with limited data remains a significant challenge.

\subsection{Insight}
When a kernel runs on a GPU, its dynamic power primarily stems from two sources: computation and memory access~\cite{kandiah2021accelwattch}. The computational power is largely determined by the number of integer and floating-point operations performed by the kernel. Additionally, the energy consumed during memory access is chiefly influenced by the number of accesses to various cache levels. In the context of deep learning, where most kernels are data-parallel, these kernels can often be represented through loops. By analyzing the depth of these loops and the innermost non-loop statements, one can extract features related to the kernel's compute and memory access volumes~\cite{zheng2020ansor}. These features could then be used to predict the kernel’s energy consumption.

\subsection{Approach}
To enhance the speed and accuracy of kernel energy prediction, we have employed a machine-learning-based cost model. The foundation of our cost model is built upon the XGBoost model used in previous works~\cite{chen2018tvm,zheng2020ansor}, which is a scalable tree boosting algorithm extensively utilized in data science. Structurally, XGBoost constructs an ensemble of decision trees in a sequential manner, with the model's predictions derived from the cumulative output of all trees.

To ensure our cost model accurately predicts the energy of a kernel, we extract high-level features related to each kernel's arithmetic operations and memory access. These features include the number of floating-point and integer operations, vectorization-related features, loop-related features, and cache access features. By incorporating these features as inputs into the cost model, it predicts a normalized energy score for the kernel.
During the training of our cost model, we employ a weighted squared error as the loss function in XGBoost. The mathematical expression for the loss function is given below, where $E_p$ and $E_m$ represent the predicted energy and the measured energy, respectively. This loss function assigns higher training weights ($\frac{1}{E_m}$) to kernels with lower energy, thereby enhancing the model’s accuracy in predicting the energy consumption of such kernels.
\begin{equation}
Loss(E_p, E_m) = \frac{(E_p-E_m)^2}{E_m}
\end{equation}

\section{Dynamic energy cost model updating strategy for fast compilation} ~\label{sec:strategy}
In this section, we introduce our energy cost model updating strategy, which aims to reduce the need for extensive energy measurements and accelerate the search process.

\subsection{Motivation}
In our energy-aware kernel generation method, illustrated in Figure~\ref{fig:overview}, each genetic iteration involves evaluating the energy consumption of kernels. Typically, identifying the most energy-efficient kernels requires numerous evaluations, potentially reaching thousands. Relying solely on measurement tools such as the NVML API would extend the search process to several hours. To expedite the search, it is crucial to integrate our energy cost model with the search process.

\subsection{Challenge}
During the search process, genetic algorithms often produce numerous kernels with new and unique features. If a cost model is pre-trained offline using existing kernel data before the search begins, the data distribution of these new kernel features might differ significantly from that of the cost model. This discrepancy could lead to inaccurate energy evaluations when using the cost model during the search, potentially affecting the quality of the generated kernels.

\subsection{Insight}
To strike a balance between accurate energy evaluation and fast compilation, we blend NVML-based energy measurements of kernels with estimated energy assessments from the cost model. To accomplish this, we devise a strategy for dynamically updating the energy cost model based on prediction errors. When the prediction error of the cost model is low, the system requires fewer kernels for energy measurements. Conversely, if the prediction error is high, more kernel measurements are necessary.

\begin{algorithm}
\footnotesize
\begin{flushleft}
  \SetAlgoLined
        \red{\# Reproduce a new kernel generation with parent kernels using genetic algorithm} \\
        Kernel\_generation $\leftarrow$ GeneticReproduction(Kernel\_parents) \\
        \red{\# Get the latency of kernels and pick the fastest M ones} \\
        Kernel\_M $\leftarrow$ LatencyEvaAndPick(Kernel\_generation, M) \\
        \red{\# Evaluate the M kernels with Energy Cost Model and pick the most energy efficient k*M kernels and their predicted energy} \\
        Kernel\_kM, EnergyPredicted\_kM $\leftarrow$ EnergyModelEvaAndPick(Kernel\_M, EnergyModel, k*M) \\
        \red{\# Get the NVML measured energy of the k*M kernels} \\
        EnergyMeasured\_kM $\leftarrow$ NVMLMeasurement(Kernel\_kM) \\
        \red{\# Update the energy cost model with measured kernels} \\
        EnergyModel $\leftarrow$ ModelUpdate(EnergyModel, EnergyMeasured\_kM) \\
        \red{\# Calculate the signal-to-noise ratio (SNR) error of the energy cost model prediction} \\
        PredictionError $\leftarrow$ SNR(EnergyPredicted\_kM, EnergyMeasured\_kM) \\
        
        \red{\# Update k vaule according to the prediction error} \\
        \If{PredictionError < $\mu$}{
          k $\leftarrow$ Min(1.0, k + 0.2)
        }
        \Else{
          k $\leftarrow$ Max(0.0, k - 0.2)
        }
        
        \red{\# Select top 50\% lower energy kernels for the next round} \\
        Kernel\_parents $\leftarrow$ EnergyModelEvaAndPick(Kernel\_M, EnergyModel, 0.5*M)
        
    \caption{Each searching round after initial round in our compilation with dynamic energy evaluation strategy. Such round will end until the kernels searched converge.}
    \label{alg:pipeline}
\end{flushleft}
\end{algorithm}

\subsection{Approach}
In our approach, we continuously monitor the predictive accuracy of the cost model. If the accuracy surpasses a predetermined threshold, it suggests that the current kernel's feature distribution aligns with the cost model's training data distribution. In such cases, a small subset of empirically measured energy data from kernels suffices to update the cost model. Conversely, if the accuracy falls below the threshold, we opt to update the model using a larger set of kernel data. We will introduce our method in the following sections.

At the beginning of searching, we randomly generate numerous kernels and identify M kernels with better latency after latency evaluation. Their energy data, gathered via NVML, is used to train the cost model for the first time. 
In each subsequent round shown in Algorithm~\ref{alg:pipeline}, the genetic algorithm and latency evaluation first yield M top-latency kernels. 
Then, our energy cost model estimates their energy and selects top k*M energy-efficient kernels with 'k' starting at 1.0 value. 
At this point, NVML is used to measure the energy of these k*M kernels, and the results are compared against the cost model's predictions.
An prediction error below the preset threshold $\mu$ implies that the cost model requires only minor updates. 
Consequently, after updating the cost model with k*M kernels, the value of k decreases for the next iteration of search. If the prediction error exceeds $\mu$, on the other hand, the value of k will increase after updating the model for a broader set of kernel data in the subsequent rounds.
At the end of each search round, the parent kernels with lower energy are selected for reproduction in the next round.

Incorporating the energy cost model into our framework can significantly accelerate the kernel generation process. For instance, if the value of k drops to 0.5 during a search round, only M/2 kernels need to be measured for updating the cost model, compared to M measurements if the cost model were not utilized in the search process. This results in nearly a 2x speed-up, as energy measurement typically consumes the most time in a search round. As the process progresses, the number of kernels requiring measurement can be further reduced after several rounds, thereby achieving an even higher acceleration ratio.

\section{Experiments} ~\label{sec:exp}

In this section, we present the experimental evaluation results of our methods. We begin with an overview of the experimental setups. Following that, we provide the energy consumption and latency results of the kernels discovered through our search process. Finally, we discuss the search speed of our system and evaluate the prediction efficiency of the energy cost model.

\subsection{Experimental setup} 
In the experiments, we focus on the operators from transformer-based and convolution-based models, including general matrix multiplication (MM) operators, matrix and vector multiplication (MV) operators, and convolution (Conv) operators. The shapes of MM, MV, and Conv are presented in the format of (batch size, M, N, K), (batch size, M, N, K), and (batch size, height, width, input channel, output channel, kernel size, stride, padding), respectively. Our experimental platform includes two types of GPUs: an NVIDIA A100 GPU (Ampere architecture) and an NVIDIA RTX 4090 GPU (Ada Lovelace architecture). To show the effectiveness of our method, we select the state-of-the-art open-source auto-scheduler Ansor~\cite{zheng2020ansor} as the baseline and implement our compilation framework based on it. All experiments are conducted using the FP32 format.

\begin{table*}[]
\caption{The energy reduction and latency impact on MM, MV, and CONV operators, running on NVIDIA A100.}
\resizebox{2\columnwidth}{!}{%
\begin{tabular}{c|cccc|cccc|ccc|c}
\hline
Energy (mJ) & MM1 & MM2 & MM3 & MM4 & MV1 & MV2 & MV3 & MV4 & CONV1 & CONV2 & CONV3 & \textbf{Average} \\ \hline
Ansor & 8.3 & 47.23 & 56.09 & 375.18 & 494.06 & 434.72 & 29.17 & 109.14  & 68.47 & 89.47 & 324.37 &  \\
Ours & 6.5 & 45.07 & 54.36 & 325.94 & 479.65 & 427.27 & 27.85 & 106.51  & 59.16 & 77.79 & 319.39 &  \\ \hline
Energy reduction (\%) & 21.69\% & 4.57\% & 3.08\% & 13.12\% & 2.92\% & 1.71\% & 4.53\% & 2.41\%  & 13.60\% & 13.05\% & 1.54\% & \textbf{7.47\%} \\ \hline
Latency (ms) & MM1 & MM2 & MM3 & MM4 & MV1 & MV2 & MV3 & MV4 & CONV1 & CONV2 & CONV3 & \textbf{Average} \\ \hline
Ansor & 0.0347 & 0.147 & 0.197 & 1.249 & 1.532 & 1.42 & 0.121 & 0.399  & 0.326 & 0.253 & 0.973 &  \\
Ours & 0.0352 & 0.154 & 0.178 & 1.243 & 1.559 & 1.412 & 0.106 & 0.369  & 0.337 & 0.263 & 0.973 &  \\ \hline
Latency increased (\%) & 1.44\% & 4.76\% & -9.64\% & -0.48\% & 1.76\% & -0.56\% & -12.40\% & -7.52\% & 3.37\% & 3.95\% & 0.00\%  & \textbf{-1.39\%} \\ \hline
\end{tabular}%
}
\label{tab:A100ansor}
\end{table*}

\subsection{Evaluation on kernel energy and latency}
\textbf{Results on NVIDIA A100 GPU.}
To show the effectiveness of our method on NVIDIA A100, we conduct comparison experiments on both computation-intensive (MM, Conv) and memory-access-intensive (MV) operators, following~\cite{zheng2020ansor}.
For each type of operator, the experiments cover various operator shapes and batch sizes to verify the effectiveness of our method under different settings.
Specifically, MM and MV shapes include MM1(1, 512, 512, 512), MM2(1, 1024, 1024, 1024), MM3(8, 512, 512, 512), MM4(8, 1024, 1024, 1024), MV1(1, 1, 49512, 12288), MV2(1, 1, 32768, 16384), MV3(8, 1, 4096, 1024), and MV4(8, 1, 8192, 2048). The shapes of CONV include CONV1(8, 7, 7, 512, 512, 3, 1, 1), CONV2(16, 56, 56, 64, 64, 1, 1, 0), and CONV3(64, 56, 56, 64, 64, 1, 1, 0). 

The results are shown in Table~\ref{tab:A100ansor}. We find our method is effective for all types of operators compared with our baseline, Ansor. The energy reduction rate varies among all settings, depending on the shape of operators and the optimal memory access ratio for each operator. The largest energy reduction is 21.69\%, occurring in MM1. We will analyze the reason for this case and present the case study in Section~\ref{sec:casestudy}. On average, our method reduces energy by 7.47\%, which is considered significant because we only involve kernel-level optimization in this work. Moreover, the power reduction is substantial in some cases. For example, the operating power of MM1 drops from 239W with Ansor to 184W, which could lead to significant energy cost reductions in GPU cluster cooling systems. Regarding kernel latency, all evaluated kernels maintain similar latency compared to those generated by Ansor. This is because our method prioritizes kernel latency during the search, selecting kernels with better energy efficiency from those with lower latency in each genetic round. It is worth noting that introducing energy considerations during the search process can even lead to finding kernels with improved latency in some cases (e.g., MM3, MV3). We believe this is an interesting topic worth studying further in the future.

\begin{figure*}[]  
    \centering  
        \includegraphics[width=0.85\textwidth]{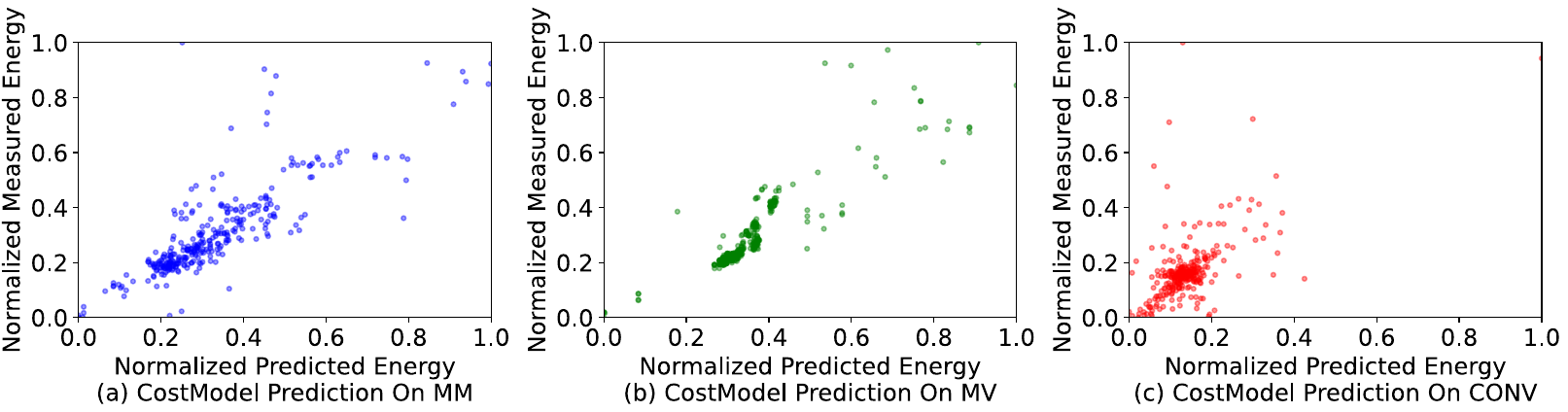}  
    \caption{The normalized predicted energy v.s. the normalized measured energy.} 
    \label{fig:costmodel} 
\end{figure*}

\textbf{Results on NVIDIA RTX 4090 GPU.}
To show the effectiveness of our method regardless of the hardware platform, we conduct further experiments on the NVIDIA RTX 4090 GPU. The operators evaluated include MM(1, 512, 512, 512), MV(1, 1, 4096, 1024), and CONV(16, 56, 56, 64, 64, 1, 1, 0), as shown in Table~\ref{tab:4090}. The conclusions are similar to those on the NVIDIA A100, with kernels achieving lower energy consumption and limited latency impact. Notably, the kernel energy reduction is even higher on the RTX 4090. For example, the MV operator consumes more than 50\% less energy compared to when using Ansor. Considering that MV operators are widely used in the deployment of large language models, this significant energy reduction could lead to substantial savings for companies.

\begin{table}[h]
\caption{The energy reduction and latency impact on MM, MV, and CONV operators, running on NVIDIA RTX 4090.}
\resizebox{0.8\columnwidth}{!}{%
\begin{tabular}{c|ccc}
\hline
Energy (mJ) & MM & MV & CONV \\ \hline
Ansor & 3.77 & 6.909 & 39.41 \\
Ours & 3.32 & 3.238 & 31.85 \\ \hline
Energy reduction (\%) & 11.94\% & 53.13\% & 19.18\% \\ \hline
Latency (ms) & MM & MV & CONV \\ \hline
Ansor & 0.0126 & 0.0118 & 0.0842 \\
Ours & 0.0137 & 0.0123 & 0.0918 \\ \hline
Latency increased (\%) & 8.73\% & 4.24\% & 9.03\% \\ \hline
\end{tabular}%
}
\label{tab:4090}
\end{table}

\begin{table}[h]
\caption{The energy and latency comparisons between our method and cuBLAS.}
\resizebox{0.8\columnwidth}{!}{%
\begin{tabular}{c|cccc}
\hline
Energy (mJ) & MM1 & MM2 & MV1 & MV2 \\ \hline
cuBLAS & 7.19 & 51.43 & 481.57 & 424.82 \\
Ours & 6.5 & 45.07 & 479.65 & 427.27 \\ \hline
Latency (ms) & MM1 & MM2 & MV1 & MV2 \\ \hline
cuBLAS & 0.0308 & 0.140 & 1.421 & 1.266 \\
Ours & 0.0352 & 0.154 & 1.559 & 1.412 \\ \hline
\end{tabular}%
}
\label{tab:cublas}
\end{table}

\textbf{Comparisons kernels from manual library cuBLAS. }
cuBLAS (CUDA Basic Linear Algebra Subprograms) is a GPU-accelerated library for linear algebra operations developed by NVIDIA. In Table~\ref{tab:cublas}, we show the comparison between kernels generated by our method and those from cuBLAS. Operators under test involve MM1(1, 512, 512, 512), MM2(1, 1024, 1024, 1024), MV1(1, 1, 49512, 12288), and MV2(1, 1, 32768, 16384). Despite the fact that manual GPU kernels often have better quality than automatically searched kernels, our kernels still achieve considerable energy reduction. For example, the energy reduction on MM1 can reach nearly 10\%. In terms of latency, cuBLAS kernels demonstrate their superiority, a finding that has been reported in previous works~\cite{chen2018tvm,zheng2020ansor}. We believe this gap can be narrowed if we use manual kernels as the initial population at the beginning of the searching process. We leave this as future work.

\subsection{Evaluation on energy cost model}

To show the prediction quality of our energy cost model, we present the ratio of normalized measured energy to normalized predicted energy in Figure~\ref{fig:costmodel}. The evaluated operators include MM(1, 512, 512, 512), MV(1, 1, 4096, 1024), and CONV(16, 56, 56, 64, 64, 1, 1, 0). We collected thousands of kernel energy data points, dividing them into training data (80\%) and test data (20\%). In these three sets of experiments, the normalized measured energy and the normalized predicted energy demonstrate a strong linear relationship. This indicates that our energy cost model can estimate the relative energy consumption of different kernels with reasonable accuracy, thereby enhancing the efficiency of kernel reproduction.

\subsection{Evaluation on searching speed}
\begin{figure}[]  
    \centering  
        \includegraphics[width=0.33\textwidth]{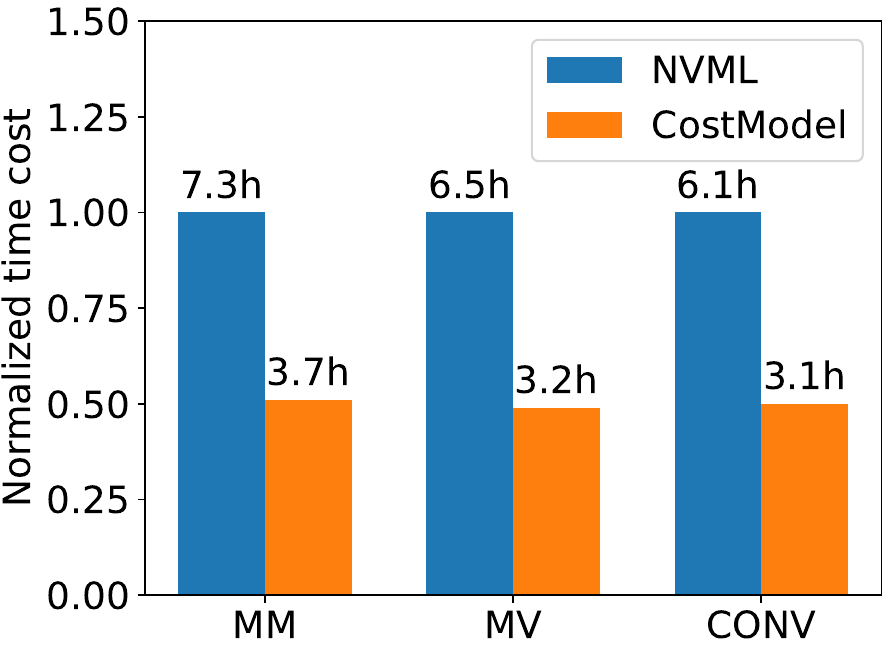}  
    \caption{The time cost of NVML-only and costmodel-based searching.} 
    \label{fig:speedup} 
\end{figure}

Since the cost model predicts kernel times in milliseconds, unlike the NVML-only method which requires several seconds, integrating the cost model into the search process can significantly reduce energy measurement time, thereby enhancing search speed. In Figure~\ref{fig:speedup}, we compare the speedup achieved by the cost model-based method versus the NVML-only method using the same test operators as in Table~\ref{tab:4090}. We generated 1000 kernels on the NVIDIA A100 with both methods and adjusted the $\mu$ value (in Algorithm~\ref{alg:pipeline}) to nearly halve the number of NVML measurements. The results show that incorporating the cost model makes the system nearly twice as fast as the NVML-only method, validating the effectiveness of our cost model's dynamic updating strategy.

\section{Case study} \label{sec:casestudy}
We examine a representative case to demonstrate why our method reduces the energy consumption of identified kernels while maintaining competitive latency. We selected an MM operator where the computation dimensions M, N, and K are all set to 512, and conducted our tests on an NVIDIA A100 GPU. The profiling data is presented in Table~\ref{tab:profiling}. We primarily attribute the energy difference to variations in static energy and memory access energy.

\begin{table}[h]
\caption{The profiling data of our kernel (K1, latency 0.0352 ms, energy 6.5 mJ) and the kernel by Ansor (K2, latency 0.0347ms, energy 8.3 mJ).}
\resizebox{\columnwidth}{!}{%
\begin{tabular}{c|ccccccc}
\hline
 & grid & block & sm\_efficiency & glb\_ld & glb\_st & shared\_ld & shared\_st \\ \hline
K1 & 64 & 256 & 55.95\% & 524288 & 131072 & 1572864 & 131072 \\ 
K2 & 256 & 128 & 83.31\% & 1310720 & 32768 & 2621440 & 327680 \\\hline
\end{tabular}%
}
\label{tab:profiling}
\end{table}

A key factor in reducing static energy is the difference in kernel grid sizes, which determine the number of blocks assigned to each GPU Streaming Multiprocessor (SM). K1 has a grid size of 64, compared to K2's 256. With the A100 having 108 SMs, K1's smaller grid size means fewer SMs are active during execution. This leads to several SMs idling, reducing SM efficiency from 83.31\% in K2 to 55.95\% in K1. More idle SMs can lead to lower static energy consumption.
The variation in memory access energy results from differences in the kernels' block sizes, which refer to the number of threads per block. K1's block size is 256, which is twice that of K2. This larger block size implies increased data reuse within each block, resulting in fewer global (glb\_ld) and shared memory load transactions (shared\_ld). Consequently, this reduces memory access energy consumption.

In terms of latency, although K1 operates with fewer SMs, the enhanced data reuse within each SM boosts computational efficiency per SM. In contrast, while K2 engages more SMs, they are not fully utilized, leading to inefficiencies. As a result, K1 and K2 achieve similar latency.
\section{Conclusion} ~\label{sec:conclusion}
This paper focuses on reducing energy consumption for deep learning kernels on GPUs. We propose a novel search-based compilation method that incorporates energy metrics into the search process to generate energy-efficient GPU kernels. Additionally, we develop a fast and accurate energy cost model to expedite the energy evaluation process. To minimize the number of energy measurements required, we implement a dynamic updating strategy for the energy cost model, further speeding up the search process. Our evaluation demonstrates that this approach can produce GPU kernels with up to 21.69\% reduced energy consumption, while maintaining good latency performance.

\bibliographystyle{ACM-Reference-Format}
\bibliography{sample-base}

\end{document}